\documentclass[12pt]{article}

\usepackage{amssymb}
\usepackage{amsmath}
\usepackage{amscd}
\usepackage{latexsym}
\usepackage{graphicx}

\usepackage{cite}
\usepackage{enumitem}

\newcommand{\be}{\begin{equation}}
\newcommand{\ee}{\end{equation}}
\newcommand{\Dlt}{\Delta}
\newcommand{\dlt}{\delta}
\newcommand{\prt}{\partial}
\newcommand{\br}{{\bf r}}
\newcommand{\bk}{{\bf k}}
\newcommand{\bfe}{{\bf e}}
\newcommand{\bn}{{\bf n}}
\newcommand{\bP}{{\bf P}}
\newcommand{\bB}{{\bf B}}
\newcommand{\bM}{{\bf M}}
\newcommand{\bE}{{\bf E}}
\newcommand{\bS}{{\bf S}}
\newcommand{\bD}{{\bf D}}
\newcommand{\bH}{{\bf H}}
\newcommand{\bJ}{{\bf J}}
\newcommand{\bt}{\beta}
\newcommand{\vp}{\varphi}

\newcommand{\al}{\alpha}
\newcommand{\sgm}{\sigma}

\newcommand{\gm}{\gamma}
\newcommand{\om}{\omega}
\newcommand{\Om}{\Omega}

\newcommand{\lbd}{\lambda}

\newcommand{\rgl}{\rangle}
\newcommand{\lgl}{\langle}

\begin{document}

\begin{center}

{\Large{\bf Superradiance by ferroelectrics in cavity resonators} \\ [5mm]

V.I. Yukalov } \\ [3mm]

{\it Bogolubov Laboratory of Theoretical Physics, \\
Joint Institute for Nuclear Research, Dubna 141980, Russia \\ [2mm]
and \\ [2mm]
Instituto de Fisica de S\~ao Carlos, Universidade de S\~ao Paulo, \\
CP 369,  S\~ao Carlos 13560-970, S\~ao Paulo, Brazil  } \\ [5mm]
{\bf e-mail}: yukalov@theor.jinr.ru
\end{center}

\vskip 2cm

\begin{abstract}

A theory is presented showing that, under appropriate conditions, a ferroelectric in 
a cavity resonator can emit superradiant pulses. Initially, the ferroelectric has to 
be prepared in a nonequilibrium state from which it relaxes emitting a coherent pulse 
in the infrared region. Polarization dipolar waves play the role of the triggering 
mechanism initiating the beginning of the process.   

\end{abstract}

\vskip 1cm

{\parindent=0pt
{\bf Keywords}: superradiance, cavity resonators, Purcell effect, scale separation
  }

\newpage

\section{Introduction}

An ensemble of atoms or molecules, as is known 
\cite{Allen_1,Samartsev_2,Gross_3,Naboikin_4,
Andrianov_5,Zinoviev_6,Zinoviev_7,Andreev_8,Benedict_9,Zinoviev_10,Kalachev_11}
can emit superradiant electromagnetic pulses due to atomic interactions through the 
common radiation field. This type of coherent emission, where the process starts with 
spontaneous atomic radiation, was first described by Dicke \cite{Dicke_12} and is often 
called {\it Dicke superradiance} \cite{Yukalov_13}. The same mathematical description, 
as for atomic superradiance, is valid for exciton superradiance 
\cite{Kopvillem_14,Bashkirov_15,Aaviksoo_16,Andrianov_17,
Andrianov_18,Hanamura_19,Sazonov_20,Itoh_21,Boer_22,Fidder_23,Devead_24,Knoester_25,
Bjork_26,Tokihiro_27,Bjork_28,Wang_29,Agranovich_30,Chen_31,Chen_32,Jiu_33}, including 
superradiance from quantum dots and wells 
\cite{Singh_34,Chen_35,Temnov_36,Parascandolo_37,Schneiber_38,Sitek_39,Yukalov_40} and 
for polariton superradiance 
\cite{John_41,Malyshev_42,Yukalov_43,Yukalov_44,Yukalov_45,Yukalov_46}. 

There also exists superradiance of non-Dicke type, when radiating dipoles cannot 
be collectivized through the common radiation field, but can become correlated by 
a resonator feedback field, which is called {\it Purcell effect} \cite{Purcell_47}. 
This type of superradiance, due to the Purcell effect, is typical of superradiance 
produced by spin and quasi-spin assemblies composed, e.g., of polarized nuclei, 
magnetic nanomolecules, magnetic nanoclusters, dipolar atoms and molecules, spinor 
atoms and molecules, and ferromagnets. Numerous citations can be found in review 
articles \cite{Yukalov_48,Yukalov_49} and recent publications
\cite{Yukalov_50,Yukalov_51,Yukalov_52,Yukalov_53}.     
   
In the present paper, we show that ferroelectrics can also produce superradiance. 
Similarly to spin systems, ferroelectric superradiance is possible only when the 
ferroelectric sample is connected to a resonator, so that the ferroelectric 
superradiance also is of non-Dicke type. A kind of luminescence can be emitted by 
polarized samples not coupled to a resonator \cite{Patel_54}, but this radiation is 
not coherent. The possibility of coherent radiation by ferroelectrics was mentioned 
in \cite{Yukalov_55}, but the full theory was not developed. Here the theory of 
ferroelectric superradiance is given with all details that are necessary for 
proving the feasibility of this phenomenon.  

Throughout the paper, the system of units is used, where the Planck and Boltzmann 
constants are set to one.

\section{Ferroelectric model}

Let us consider a ferroelectric of the order-disorder type, whose Hamiltonian 
is \cite{Blinc_56}
\be
\label{1}
 \hat H = -\Om \sum_j S_j^x \; - \;\frac{1}{2} \sum_{i\neq j} J_{ij} S_i^z S_j^z \;
- \; \sum_j \bE_{tot} \cdot \hat\bP_j \;  .
\ee
Here $\Omega$ is tunneling frequency; $J_{ij}$ is the interaction potential between 
two lattice sites enumerated by the indices $i, j = 1,2,\ldots,N$, where the 
self-action is excluded through the condition $J_{jj} = 0$. The site polarization 
operator, as is known \cite{Blinc_56}, depends on the summetry of the potential well 
at the lattice site. If at the lattice site there is a symmetric double-well, then 
the polarization operator contains only the $z$-component. However, we consider the 
general case of a nonsymmetric potential at lattice sites, when the polarization 
operator has the form 
\be
\label{2}
 \hat\bP_j = d_0 \bS_j
\ee
expressed through the electric dipole $d_0$ and the spin one half operators with 
the commutation relations
$$
[S_i^x , \; S_j^y ] = i\dlt_{ij} S_j^z \; , \qquad  
[S_i^y , \; S_j^z ] = i\dlt_{ij} S_j^x \; , \qquad 
[S_i^z , \; S_j^x ] = i\dlt_{ij} S_j^y \; .
$$
The total electric field 
\be
\label{3} 
\bE_{tot} = E \bfe_x + E_0 \bfe_z
\ee
consists of a resonator feedback field $E$ and an external field $E_0$.  

The Heisenberg equations of motion for the spin operators yield
$$
\frac{dS_i^x}{dt} = \left( \sum_j J_{ij} S_j^z + d_0 E_0  \right) S_i^y \; ,
$$
$$
\frac{dS_i^y}{dt} = - \left( \sum_j J_{ij} S_j^z + d_0 E_0  \right) S_i^x +
( \Om + d_0 E ) S_i^z \; ,
$$
\be
\label{4}
\frac{dS_i^z}{dt} = - ( \Om + d_0 E ) S_i^y \; .
\ee
Our aim is to find the temporal behaviour of the averages
\be
\label{5}
  s_\al \equiv \frac{2}{N} \sum_j \lgl \; S_j^\al \; \rgl \; .
\ee
For the average interaction potential, we shall use the notation
\be
\label{6}
J \equiv \frac{1}{N} \sum_{i\neq j} J_{ij} \; .
\ee
To take into account spin relaxation, we employ the method of local fields 
\cite{Wangsness_57,Yukalov_58}, where particle interactions are considered as acting 
in the local field formed by other particles so that there appears the attenuation of 
spin motion, which forces the spin variables at each moment of time to relax to their 
locally-equilibrium values. The latter are defined as having the form of the 
equilibrium averages
\be
\label{7}
 \zeta_\al \equiv \frac{2}{N} \sum_j \lgl \; S_j^{\al} \; \rgl_{eq} \; ,
\ee
but expressed through the variables (\ref{5}) taken at the given moment of time. 

The analysis of the evolution equations can be done by invoking {\it scale separation 
approach} 
\cite{Yukalov_48,Yukalov_49,Yukalov_50,Yukalov_51,Yukalov_52,Yukalov_53,Yukalov_55}.   
The pair spin correlators are decoupled by means of the stochastic mean-field 
approximation \cite{Yukalov_48,Yukalov_49,Yukalov_50} giving
\be
\label{8}
 \lgl \; S_i^\al S_j^\bt \; \rgl = \lgl \; S_i^\al \; \rgl \lgl \; S_j^\bt \; \rgl
+ \lgl \; S_i^\al \; \rgl \dlt S_j^\bt + \lgl \; S_j^\bt \; \rgl \dlt S_i^\al \; ,
\ee
where $i \neq j$ and $\delta S_j^\alpha$ are treated as stochastic variables with zero 
mean,
\be
\label{9}
 \lgl \lgl \; \dlt S_j^\al \; \rgl \rgl = 0 \;  .
\ee
This approximation makes it possible to take into account spin correlations caused 
by spin waves, which is principally important at the initial stage of spin relaxation.  

Averaging equations (\ref{4}), with decoupling (\ref{8}), we meet the stochastic 
variable
\be
\label{10}
 \xi_i^\al \equiv \sum_j J_{ij} \dlt S_j^\al \;  .
\ee
Under the averaging over the sample, we use the mean-field type approximation
$$
\frac{1}{N} \sum_j \xi_j^z \lgl \; S_j^\al \; \rgl = 
\xi_0 \;\frac{1}{N} \sum_j \; \lgl \; S_j^\al \; \rgl \; ,
$$
\be
\label{11}
 \frac{1}{N} \sum_j \xi_j^\al \lgl \; S_j^z \; \rgl = 
\xi_\al \;\frac{1}{N} \sum_j \; \lgl \; S_j^z \; \rgl \; ,
\ee
in which $\xi_0$ and $\xi_\alpha$ are stochastic variables. Thus we come to the 
equations
$$
\frac{ds_x}{dt} = \left( \frac{J}{2}\; s_z + d_0 E_0 + \xi_0 \right) s_y + 
\xi_y s_z - \gm_2 ( s_x - \zeta_x ) \; ,
$$
$$
\frac{ds_y}{dt} = - \left( \frac{J}{2}\; s_z + d_0 E_0 + \xi_0 \right) s_x + 
( \Om + d_0 E - \xi_x ) s_z - \gm_2 ( s_y - \zeta_y ) \; ,
$$
\be
\label{12}
\frac{ds_z}{dt} = - ( \Om + d_0 E ) s_y - \gm_1 ( s_z - \zeta_z ) \; .
\ee

It is also useful to consider the ladder operator
$$
 S_j^\pm \equiv S_j^x \pm i S_j^y \;  ,
$$
whose average yields the variable
\be
\label{13}
 u \equiv \frac{2}{N} \sum_j \lgl \; S_j^- \; \rgl = s_x - i s_y \;  .
\ee
Then, denoting the quantities
\be
\label{14}
 \xi \equiv \xi_x - i \xi_y \; , \qquad \zeta \equiv \zeta_x - i \zeta_y \; ,
\ee
we obtain the equation
\be
\label{15}
 \frac{du}{dt} = i \left( \frac{J}{2}\; s_z + d_0 E_0 + \xi_0 \right) u -
i (\Om + d_0 E - \xi ) s_z - \gm_2 ( u - \zeta ) \; .
\ee

The stochastic variables are assumed to satisfy the correlation conditions
$$
\lgl \lgl \;\xi_0(t) \; \rgl \rgl =  \lgl \lgl \;\xi(t) \; \rgl \rgl = 0\; ,
$$
\be
\label{16}
\lgl \lgl \;\xi_0(t) \xi(t') \; \rgl \rgl = 0\; ,  \qquad
\lgl \lgl \;\xi_0(t) \xi_0(t') \; \rgl \rgl = 
\lgl \lgl \;\xi^*(t) \xi(t') \; \rgl \rgl = 2\gm_3 \dlt( t - t' ) \; ,
\ee
where stochastic averaging is implied.

\section{Locally-equilibrium state}

In order to define the locally-equilibrium (quasi-equilibrium) values (\ref{7}), it 
is necessary to consider a Hamiltonian without the resonator feedback field, 
\be
\label{17}
\hat H_{eq} = - \Om \sum_j S_j^x \; - \; 
\frac{1}{2} \sum_{i\neq j} J_{ij} S_i^z S_j^z \; - \;
d_0 E_0 \sum_j S_j^z \; .
\ee
Resorting to the mean-field approximation gives
\be
\label{18}
\hat H_{eq} = - \Om \sum_j S_j^x \; - \; J_{eff} \sum_j S_j^z \;   ,
\ee
where
\be
\label{19}
 J_{eff} \equiv J \lgl \; S_j^z \; \rgl + d_0 E_0 \;  .
\ee
Calculating statistical averages of spin components at temperature $T$, we employ the 
notation
\be
\label{20}
\Om_{eff} \equiv \sqrt{\Om^2 + J_{eff}^2} \; .
\ee
Thus we obtain
$$
\lgl \; S_j^x \; \rgl_{eq} = 
\frac{\Om}{2\Om_{eff}} \; \tanh\left( \frac{\Om_{eff}}{2T} \right) \; ,
\qquad \lgl \; S_j^y \; \rgl_{eq} = 0 \; ,
$$
\be
\label{21}
\lgl \; S_j^z \; \rgl_{eq} = 
\frac{J_{eff}}{2\Om_{eff}} \; \tanh\left( \frac{\Om_{eff}}{2T} \right) \; .
\ee

The external electric field is directed down, so that 
\be
\label{22}
\om_0 \equiv - d_0 E_0 > 0 \; .
\ee
Replacing in equations (\ref{21}) the average spins by their time-dependent values at 
zero temperature yields
$$
\zeta_x = \frac{\Om}{\Om_{eff}} = 
\frac{2\Om}{[4\Om^2+(J s_z-2\om_0)^2]^{1/2}} \; , \qquad \zeta_y =  0 \; ,
$$
\be
\label{23}
\zeta_z = \frac{J_{eff}}{\Om_{eff}} = 
\frac{Js_z-2\om_0}{[4\Om^2+(J s_z-2\om_0)^2]^{1/2}} \;   .
\ee
Since $s_z = s_z(t)$ is a function of time, the locally-equilibrium quantities also 
depend on time,
$$
\zeta_\al = \zeta_\al(t) \; , \qquad \zeta = \zeta(t) = \zeta_x(t) \; .
$$

\section{Resonator field}

To derive an equation for the resonator feedback field, it is possible to follow the 
general methods for treating cavity resonators \cite{Mandel_59,Checchin_60}. Here we 
keep in mind a cylindrical resonator cavity of radius $R$, length $L$, and volume 
$V_{res} = \pi R^2 L$. The axis of the resonator is along the axis $x$. The Gaussian 
system of units will be used.

Electromagnetic fields inside the resonator obey the Maxwell equations
$$
\vec{\nabla}\cdot\bD = 4\pi\rho \; , \qquad \vec{\nabla}\cdot\bB = 0 \;
$$
\be
\label{24}
 \vec{\nabla}\times \bE = -\; \frac{1}{c} \;\frac{\prt\bB}{\prt t} \; , 
\qquad
 \vec{\nabla}\times \bH = 
\frac{4\pi}{c} \; \bJ + \frac{1}{c} \;\frac{\prt\bD}{\prt t} \;  ,
\ee
in which
\be
\label{25}
\bD = \bE + 4\pi\bP \; , \qquad  \bB = \bH + 4\pi\bM \;  .
\ee
We consider the case of a ferroelectric inserted into the resonator, where there are 
no free charges and magnetic inclusions, so that the material equations are
\be
\label{26}
 \rho = 0 \; , \qquad \bM = 0 \; , \qquad \bJ = \sgm\bE \;  .
\ee
From the Maxwell equations, we find
\be
\label{27}
 \nabla^2\bE \; -\; \frac{1}{c^2} \; \frac{\prt^2\bE}{\prt t^2} \; - \;
 \frac{4\pi\sgm}{c^2} \; \frac{\prt \bE}{\prt t} = 
\frac{4\pi}{c^2} \; \frac{\prt^2\bP}{\prt t^2} \; - \;
4\pi\vec{\nabla} (\vec{\nabla}\cdot\bP ) \; .
\ee
And the polarization vector can be represented as
\be
\label{28}
\bP = \frac{1}{V} \sum_j \lgl \; \bP_j \; \rgl = 
\frac{d_0}{V} \sum_j \lgl \; \bS_j \; \rgl \;  .
\ee

We look for the solution of equation (\ref{27}) in the form
\be
\label{29}
\bE(\br,t) = \bfe(\br) E(t) \; ,
\ee
in which ${\bf e}({\bf r})$ describes the normal resonator modes satisfying the 
Helmholtz equation
\be
\label{30}
 \left( \nabla^2 + \frac{\om^2}{c^2} \right)  \bfe(\br) = 0 
\ee
and being normalized to the resonator volume,
\be
\label{31}
 \int |\; \bfe(\br)\; |^2 d\br = V_{res} \; .
\ee
Here $\omega$ is the resonator natural frequency. Since we have chosen the resonator 
axis along the axis $x$, we take the Laplacian in the form
$$
\nabla^2 = \frac{\prt^2}{\prt r^2} + \frac{1}{r}\; \frac{\prt}{\prt r} +
\frac{1}{r^2}\; \frac{\prt^2}{\prt \vp^2} + \frac{\prt^2}{\prt x^2} \;  .
$$

The solutions to the Helmholtz equation are termed TM$_{nlm}$ modes. We are interested 
in the solution whose electric field would be along the resonator axis, such that
\be
\label{32}
 e_x(\br) \neq 0 \; , \qquad e_y(\br) = 0 \; , \qquad e_z(\br) = 0 \;  ,
\ee
except the boundary of the resonator, where
\be
\label{33}
e_x(\br) \left\vert_{r=R} \right. = 0 \;   .
\ee
The corresponding solution is given by the TM$_{010}$ fundamental mode
\be
\label{34}
 e_x(\br) = C_0 J_0 \left(\frac{\om}{c}\; r\right) \;  ,
\ee
where $J_0$ is the Bessel function of the first kind. The boundary condition (\ref{33}), 
defining the first zero of the Bessel function,
\be
\label{35}
  J_0 \left(\frac{\om}{c}\; R\right) = 0 \; ,
\ee
prescribes the resonator natural frequency
\be
\label{36}
\om = 2.4048\; \frac{c}{R} \; .
\ee

The coefficient $C_0$ has to be found from the normalization condition (\ref{31}). To 
this end, we use the integral
$$
 \int_0^R J_\nu^2(kr) \; rdr = \frac{R^2}{2} \; \left[ J_\nu'(kR)\right]^2 +
\frac{1}{2}\; \left( R^2 \; - \;\frac{\nu^2}{k^2}\right) J_\nu^2 (kR) \;  ,
$$
in which
$$
J_\nu'(kR) = \frac{\nu}{kR}\; J_\nu(kR) - J_{\nu+1}(kR) \; .
$$
Keeping in mind the boundary condition (\ref{33}), we have
$$
J_0'(kR) = - J_1(kR) \qquad \left( k = \frac{\om}{c} \right) \; .
$$
Then the normalization condition (\ref{31}) results in the integral
$$
\int |\; e_x(\br)\;|^2 \; d\br = C_0^2 2\pi L \int_0^R J_0^2(kr)\; rdr =
C_0^2V_{res} J_1^2(kR) = V_{res} \;  ,
$$
where $k = \omega/c$, which gives
$$
 C_0 = J_1^{-1}\left( \frac{\om}{c}\; R \right) \; .
$$
Taking into account the natural frequency (\ref{36}) and the value
$$
 J_1(2.4048) = 0.519153 \;  ,
$$
we get 
\be
\label{37}
 C_0 = 1.926214 \;  .
\ee

Substituting form (\ref{29}) into equation (\ref{27}), multiplying it by 
\be
\label{38}
\bfe(\br) = C_0J_0 \left( \frac{\om}{c}\; r \right) \bfe_x \;  ,
\ee
and integrating over space yields
\be
\label{39}
 \frac{\prt^2 E}{\prt t^2} + 4\pi\sgm \; \frac{\prt E}{\prt t} + \om^2 E =
-\; \frac{4\pi}{V_{res}} \; 
\frac{\prt^2\bP}{\prt t^2} \cdot \int \bfe(\br)\; d\br \; ,
\ee
where the filling factor is
\be
\label{40}
 \eta_f \equiv \frac{1}{V_{res}} \int e_x(\br) \; d\br \; .
\ee
Introducing the resonator attenuation
\be
\label{41}
\gm \equiv 2\pi \sgm 
\ee
results in the equation
\be
\label{42}
 \frac{\prt^2 E}{\prt t^2} + 2\gm \; \frac{\prt E}{\prt t} + \om^2 E = 
-4\pi\eta_f \;  \frac{\prt^2 P_x}{\prt t^2}  \; ,
\ee
in which
\be
\label{43}
 P_x = \frac{d_0}{V} \sum_j \lgl \; S_j^x \; \rgl\;  .
\ee

Because of the integral
$$
\int_0^R J_0(kr) \; rdr = \frac{R}{k}\; J_1(kR) \; ,
$$
we get
$$
\int e_x(\br)\; d\br = 0.83167 \; V_{res} \; ,
$$
which gives
\be
\label{44}
\eta_f = 0.83167 \; .
\ee
And polarization (\ref{43}) can be written as
\be
\label{45}
 P_x = \frac{1}{2} \; \rho d_0 s_x \qquad 
\left( \rho \equiv \frac{N}{V} \right) \;  .
\ee
 
Under the initial conditions
\be
\label{46}
E(0) = 0 \; , \qquad \left. \frac{\prt E}{\prt t}\right|_{t=0} = 0 \; , \qquad
\left. \frac{\prt P_x}{\prt t}\right|_{t=0} = 0 \;   ,
\ee
the resonator equation (\ref{42}) can be rewritten as
\be
\label{47}
\frac{dE}{dt} + 2\gm E + \om^2 \int_0^t E(t')\; dt' = 
- 4\pi \eta_f \; \frac{\prt P_x}{\prt t} \;  .
\ee
This equation has the same form as the Kirchhoff equation for a resonant electric 
circuit, with a magnetic sample inside it. Therefore the consideration of the dipole 
dynamics in a ferroelectric can be done similarly to the study of spin dynamics in 
magnets \cite{Yukalov_48,Yukalov_49,Yukalov_50,Yukalov_51,Yukalov_52,Yukalov_53}.

\section{Evolution equations}

The quantity
\be
\label{48}
\om_s \equiv \om_0 \; - \; \frac{J}{2}\; s_z 
\ee
plays the role of an effective rotation frequency. Using the equality 
$$
 s_y = -\; \frac{i}{2}\; (u^* - u ) \;  ,
$$
equations (\ref{12}) can be represented as
$$
\frac{ds_x}{dt} = - ( \om_s - \xi_0 ) s_y + \xi_y s_z - \gm_2 ( s_x - \zeta_x) \; ,
$$
$$
\frac{ds_y}{dt} = ( \om_s - \xi_0 ) s_x + ( \Om + d_0 E - \xi_x ) s_z - 
 \gm_2 s_y \; ,
$$
\be
\label{49}
 \frac{ds_z}{dt} = \frac{i}{2} \; ( \Om + d_0 E  ) (u^* - u ) -  
\gm_1 ( s_z - \zeta_z) \;  .
\ee
Equation (\ref{15}) takes the form
\be
\label{50}
\frac{du}{dt} = - i (\om_s - \xi_0 - i \gm_2 ) u - i (\Om + d_0 E - \xi ) s_z +
\gm_2 \zeta \;  .
\ee
In addition, we shall need to consider the temporal behaviour of the coherence 
intensity
\be
\label{51}
 w  \equiv \frac{4}{N(N-1)} \sum_{i\neq j} \lgl \; S_i^+ S_j^- \;\rgl =
|\; u \; |^2 \;  ,
\ee
whose evolution equation reads as
\be
\label{52}
 \frac{dw}{dt} = - 2\gm_2 w - i ( \Om + d_0 E - \xi ) s_z u^* +
i ( \Om + d_0 E -\xi^* ) s_z u + \gm_2 \zeta ( u^* + u ) \; .
\ee

We assume that the detuning from the resonance is small,
\be
\label{53}
 \left| \; \frac{\Dlt}{\om} \; \right|  \ll 1 \qquad 
( \Dlt \equiv \om - \om_0 ) \; .
\ee
It is possible to take a sufficiently large external electric field, such that
\be
\label{54}
 \left| \; \frac{J}{\om_0} \; \right|  \ll 1 \;  .
\ee
All attenuations are supposed to be much smaller than the frequency $\omega_0$,
\be
\label{55}
\frac{\gm}{\om} \ll 1 \; , \qquad \frac{\gm_1}{\om_0} \ll 1 \; , \qquad
\frac{\gm_2}{\om_0} \ll 1 \;  .
\ee

As follows from equation (\ref{42}), the effective coupling rate between the 
ferroelectric sample and resonator is
\be
\label{56}
\gm_c \equiv \frac{\pi}{2} \;\eta_f \rho d_0^2
\ee
that is much smaller than the resonator natural frequency,
\be
\label{57}
 \left| \; \frac{\gm_c}{\om} \; \right| \ll 1  \; .
\ee
Solving the resonator equation (\ref{42}) by perturbation theory in powers of the 
coupling rate, in first order yields
\be
\label{58}
 d_0 E = i ( u X - X^* u^* ) \;  ,
\ee
with the coupling function
\be
\label{59}
 X = \gm_c \om_s \left[ \frac{1-\exp\{-i(\om-\om_s)t-\gm t\}}{\gm+i(\om-\om_s)} +
\frac{1-\exp\{-i(\om+\om_s)t-\gm t\}}{\gm-i(\om+\om_s)} \right] \;  .
\ee

If the effective detuning 
\be
\label{60}
\Dlt_s \equiv \om - \om_s = \Dlt + \frac{J}{2} \; s_z
\ee
is also small, such that
\be
\label{61}
 \left| \; \frac{\Dlt_s}{\om} \; \right| \ll 1  \;  ,
\ee
then in the coupling function (\ref{59}) it is possible to keep only the resonant part, 
obtaining
\be
\label{62}
X = \gm_c \om_s \; \frac{1-\exp(-i\Dlt_s t-\gm t)}{\gm+i\Dlt_s} \; .
\ee
 
Substituting expression (\ref{58}) into the evolution equation (\ref{50}) gives
\be
\label{63}
 \frac{du}{dt} = - i \om_{eff} u + i \xi_0 u + i \xi s_z - i \Om s_z +
\gm_2 \zeta - X^* u^* s_z \;  ,
\ee
where
\be
\label{64}
 \om_{eff} \equiv \om_s - i ( \gm_2 - X s_z ) \;  .
\ee
And equation (\ref{52}) leads to
$$
\frac{dw}{dt} = - 2\gm_2 ( 1 - \al s_z) w + i ( u^* \xi - \xi^* u ) s_z +
$$
\be
\label{65}
  +
i \Om (u - u^*) + \gm_2 \zeta ( u + u^*) - 
\left[ X^* (u^*)^2 + Xu^2 \right] s_z \; ,
\ee
with the notation
\be
\label{66}
 \al \equiv \frac{1}{2\gm_2} \; ( X^* + X ) = \frac{{\rm Re}\;X}{\gm_2} \; .
\ee
The equation for the polarization becomes
\be
\label{67}
\frac{ds_z}{dt} = - \al \gm_2 w - \gm_1( s_z - \zeta_z ) + 
\frac{i}{2} \;\Om ( u^* - u ) \; .
\ee
 
Taking account of the existing small parameters shows that the functional variable 
$u$ can be treated as fast, while the variables $w$ and $s_z$, as slow. Then averaging 
techniques are applicable \cite{Yukalov_55,Yukalov_61}. Solving equation (\ref{63}) for 
the fast variable, the slow variables are kept as integrals of motion. This results in 
the solution
$$
u = u_0 \exp\left\{ - i \om_{eff} t + i \int_0^t \xi_0(t')\; dt'\right\} +
$$
\be
\label{68}
 + i \int_0^t [ \; \xi(t') s_z - \Om s_z -i\gm_2\zeta \; ]
\exp\left\{ - i\om_{eff}(t-t') + i\int_{t'}^{t} \xi_0(t'')\; dt'' 
\right\} \; dt' \;  .
\ee
The found fast variable is to be substituted to the equations for the slow variables 
and their right-hand sides are to be averaged over time and over the stochastic 
variables. In that way, we obtain the equations for the guiding centers
$$
\frac{dw}{dt} = - 2\gm_2( 1 - \al s_z ) w + 2\gm_3 s_z^2 \; ,
$$
\be
\label{69}
 \frac{ds_z}{dt} = - \al \gm_2 w - \gm_1 ( s_z - \zeta_z ) \;  .
\ee
These equations are complimented by the initial conditions
\be
\label{70}
 w(0) = w_0 \; , \qquad s_z(0) = s_0 \;  .
\ee
Assuming that $|\Delta_s|$ is much smaller than $\gamma$ makes it possible to simplify 
the coupling function (\ref{66}), getting
\be
\label{71}
\al = \frac{g\gm^2}{\gm^2 + \Dlt_s^2} \;
\left( 1 \; - \; \frac{J}{2\om_0}\; s_z\right) \left( 1  - e^{-\gm t} \right) \;  ,
\ee
where
\be
\label{72}
g \equiv \frac{\gm_c\om_0}{\gm\gm_2} 
\ee
is a dimensionless coupling parameter. Under this condition, it is convenient to write 
the coupling function in the form
\be
\label{73}
\al = g ( 1 - As_z ) \left( 1  - e^{-\gm t} \right) \; ,
\ee
in which
\be
\label{74}
 A \equiv \frac{J}{2\om_0} \;  .
\ee

\section{Development of coherence}

First of all, it is necessary to emphasize that the presence of the resonator is 
crucial for organizing collective motion of dipoles. Really, the absence of the 
resonator implies that the coupling function $\al=0$. Then from equation (\ref{69}), 
it is evident that the polarization $s_z$ slowly relaxes with the relaxation rate 
$\gm_1$, while the coherence intensity $w$ slowly relaxes with the relaxation rate 
$\gm_2$, provided coherence was imposed through the initial condition. And no coherence
appears if $w_0 = 0$.

The existence of the dynamic attenuation $\gm_3$ due to dipolar waves is also of 
principal importance. If $\gm_3=0$ and no initial coherence is imposed, so that $w_0=0$, 
then coherence can never arise. The presence of $\gm_3$ initiates the motion of dipoles 
at the initial stage and leads to the arising coherence. 

Let us consider the initial stage of the process, when the coupling function is yet 
small,
\be
\label{75}
\al \simeq 0 \qquad ( \gm t \ll 1) \; .
\ee
Then equations (\ref{69}) are
\be
\label{76}
 \frac{dw}{dt} = - 2\gm_2 w + 2\gm_3 s_z^2 \; \qquad
 \frac{ds_z}{dt} = - \gm_1 (s_z - \zeta_z ) \; .
\ee
Keeping in mind that $\gamma_1 \ll \gamma$, we see that the polarization practically 
does not change,
\be
\label{77}
s_z \simeq s_0 \qquad ( \gm_1 t \ll \gm t \ll 1) \;  .
\ee
And the coherence intensity is
\be
\label{78}
 w \simeq \left( w_0 \; - \; \frac{\gm_3}{\gm_2}\; s_0^2 \right) e^{-2\gm_2 t}
+ \frac{\gm_3}{\gm_2}\; s_0^2 \; .
\ee
When either $\gamma_3 = 0$ or $s_0 = 0$, and $w_0 \neq 0$, the function $w$ slowly 
relaxes with the relaxation rate $\gamma_2$. And if $w_0 = 0$ and $s_0 \neq 0$, then
\be
\label{79}
  w \simeq \frac{\gm_3}{\gm_2}\; s_0^2 \; \left( 1 -  e^{-2\gm_2 t}\right) \qquad
( w_0 = 0 ) \;  ,
\ee
which is rather small, since usually $\gamma_3 \ll \gamma_2$. 

There is no essential coherence in the process, when $\al s_z\ll 1$. But the coupling
function $\al$ grows with time, inducing coherence. The incoherent regime lasts till
$\alpha s_z$ becomes of order of unity, after which coherence starts quickly growing. 
The time of the beginning of the coherent stage can be defined by the equality
\be
\label{80}
 \al s_z =  1 \qquad ( t = t_{coh} ) \;  ,
\ee
which yields
\be
\label{81}
 t_{coh} = \frac{1}{\gm}\; 
\ln\left[ \frac{gs_0 ( 1 - As_0 )}{gs_0(1- As_0)-1}\right] \;  .
\ee
The strength of the coupling with the resonator depends on the magnitude of the 
coupling parameter $g$ and on the initial polarization $s_0$. Under strong coupling, 
the coherence time becomes
\be
\label{82}
 t_{coh} \simeq \frac{1}{\gm g s_0(1-As_0)} \qquad ( gs_0 \gg 1 ) \; .
\ee
This also shows that coherence can develop in the sample only if the initial 
polarization corresponds to a nonequilibrium state, when $s_0$ is positive, hence 
directed against the applied electric field $E_0$. Notice that
\be
\label{83}
 \gm t_{coh} \ll 1 \qquad ( gs_0 \gg 1) \;  .
\ee
Therefore, if $\gamma_2 \ll \gamma$, then
\be
\label{84}
\gm_2 t_{coh}  \ll \gm t_{coh} \ll 1 \qquad ( gs_0 \gg 1) \;  .
\ee
Then the coherence intensity (\ref{78}) reads as
\be
\label{85}
 w \simeq w_0 + 2\gm_3 s_0^2 t \qquad ( \gm_2 t \ll 1 ) \;  .
\ee
At the coherence time, the variables $w$ and $s$ reach the values
\be
\label{86}
 w(t_{coh} ) \equiv w_{coh} \; , \qquad s(t_{coh} ) \equiv s_{coh} \;  ,
\ee
for which we have
\be
\label{87}
w_{coh} =w_0 + 2\gm_3 s_0^2 t_{coh} \; , \qquad s_{coh} = s_0 \; .
\ee
Again, let us stress the necessity for the existence of the resonator that induces 
coherence in the presence of an initial nonequilibrium polarization and the dynamic 
attenuation due to dipolar waves.

\section{Coherent stage}

Essential coherence of dipole motion develops in the sample at large $\gamma t$. Then, 
assuming that $|A| \ll 1$, one has
\be
\label{88}
\al \simeq g \qquad ( |\;A\;| \ll 1 \; , ~ \gm t \gg 1 ) \; .
\ee
Also, taking into account the standard case where
\be
\label{89}
 \gm_1 \ll g \gm_2 \; , \qquad \gm_3 \ll g\gm_2 \; ,
\ee
we get the equations
\be
\label{90}
 \frac{dw}{dt} = 2\gm_2 ( g s_z - 1) w \; , \qquad
\frac{ds_z}{dt} = -g \gm_2 w \; .
\ee
These equations enjoy exact solutions giving the coherence intensity
\be
\label{91}
w = \left( \frac{\gm_p}{g\gm_2} \right)^2 \; 
{\rm sech}^2 \left( \frac{t-t_0}{\tau_p} \right)
\ee
and polarization
\be
\label{92}
 s_z = \frac{1}{g} \; - \; 
\frac{\gm_p}{g\gm_2} \tanh \left( \frac{t-t_0}{\tau_p} \right) \; .
\ee
Here the quantities $t_0$ and $\gamma_p \equiv 1/ \tau_p$ are defined by conditions 
(\ref{86}) resulting in the equalities for the delay time
\be
\label{93}
t_0 = t_{coh} + \frac{\tau_p}{2} \; 
\ln\left| \; \frac{\gm_p+\gm_g}{\gm_p-\gm_g} \; \right|   
\ee
and pulse time
\be
\label{94}
 \tau_p \equiv \frac{1}{\gm_p} \; , \qquad
\gm_p^2 = \frac{\gm_g^2}{2} \; \left[\; 1 + 
\sqrt{ 1 + 4 \left( \frac{g\gm_2}{\gm_g}\right)^2 w_{coh} } \; \right ] \;  ,
\ee
where
\be
\label{95}
 \gm_g = ( g s_0 - 1 )\gm_2 \; .
\ee
Under strong coupling and weak initially imposed coherence, equation (\ref{94}) 
simplifies to
\be
\label{96}  
\gm_p \simeq g\gm_2 \sqrt{s_0^2 + w_{coh} } \qquad
\left( gs_0 \gg 1 \; , ~ \frac{w_{coh}}{s_0^2} \ll 1 \right) \; .
\ee
At the time $t_0$, the coherence intensity reaches its maximum, where
\be
\label{97}
w(t_0) = s_0^2 + w_{coh} \; , \qquad s_z(t_0) = \frac{1}{g} \; .
\ee
After the delay time $t_0$, the coherence intensity quickly diminishes,
\be
\label{98}
 w \simeq 4 w(t_0 ) \exp\left( - \; \frac{2}{\tau_p}\; t\right) 
\qquad ( t \gg t_0 ) \; ,
\ee
while the polarization reverses and tends to the expression
\be
\label{99}
 s_z \simeq - s_0 + \frac{2}{g} + 2s_0 \exp\left( -\; \frac{2}{\tau_p}\; t\right)
\qquad ( t \gg t_0 ) \;  .
\ee

It is useful to remark that the limit $-s_0 + 2/g$ is not an equilibrium limit, 
although the sample stays close to that state for quite a long time, slowly relaxing 
to an equilibrium value during the time $T_1 = 1/\gamma_1$. In that sense, this effect 
can be termed {\it pre-equilibration}.

\section{Radiation intensity}

The intensity of radiation by dipoles in the direction of ${\bf n} \equiv {\bf r}/|{\bf r}|$ 
at time $t$ consists of two parts, 
\be
\label{100}
 I(\bn,t) = I_{inc}(\bn,t) + I_{coh}(\bn,t) \;  ,
\ee
incoherent radiation intensity
\be
\label{101}
 I_{inc}(\bn,t) = 2\om_0 \gm_0 \sum_j \vp(\bn) \lgl \; S_j^+(t) S_j^-(t)\; \rgl  
\ee
and coherent radiation intensity
\be
\label{102}
 I_{coh}(\bn,t) = 2\om_0 \gm_0 
\sum_{i\neq j} \vp_{ij}(\bn) \lgl \; S_i^+(t) S_j^-(t)\; \rgl \; ,
\ee
in which
$$
\vp(\bn) = \frac{3}{16\pi}\; ( 1 + \cos^2\vartheta ) \; , \qquad 
\cos\vartheta = (\bn\cdot\bfe_z) \; ,
$$
$$
\vp_{ij}(\bn) = \vp(\bn) \exp\{ i k_0 \bn \cdot ( \br_i - \br_j) \} \; ,
$$
$$
\gm_0 \equiv \frac{2}{3} \; |\; {\bf d}_0 \; |^2 k_0^3 \qquad
\left( k_0 \equiv \frac{\om_0}{c} \right ) \; .
$$
The derivation of these expressions can be found in \cite{Yukalov_50}. In the present 
case, we have
$$
I_{inc}(\bn,t) = N \om_0 \gm_0 \vp(\bn) [\; 1 + s_z(t) \; ] \; ,
$$
\be
\label{103}
I_{coh}(\bn,t) = \frac{1}{2} \; N^2 \om_0\gm_0 \vp(\bn) F_0(k_0\bn) w(t) \;  ,
\ee
where
$$
F(\bk) \equiv 
\left| \; \frac{1}{N} \sum_{j=1}^N e^{i\bk\cdot\br_j}\; \right|^2 \; .
$$

A sample inside a cavity resonator can radiate only in the direction of the cavity axis, 
that is, in the direction ${\bf n} = {\bf e}_x$, when $\vartheta = \pi/2$. In this case,
$$
 \vp(\bfe_x) = \frac{3}{16\pi} \; , \qquad 
F(k_0\bfe_x) = \frac{4}{k_0^2L^2} \; \sin^2 \left( \frac{k_0L}{2}\right) \; .
$$
Then the incoherent radiation intensity is
\be
\label{104}
I_{inc}(\bfe_x,t) = \frac{3}{16\pi} \; N \om_0 \gm_0 [\; 1 + s_z(t) \; ] \; ,
\ee
while the coherent radiation intensity reads as
\be
\label{105}
 I_{coh}(\bfe_x,t) = \frac{3}{32\pi} \; N^2 \om_0 \gm_0 F(k_0\bfe_x) w(t) \; .
\ee
 
To make estimates for the radiation intensity, let us consider the system parameters 
typical of ferroelectrics \cite{Blinc_56}. The dipole interaction is $J\sim 10^2$ K 
$\sim 10^{13}$ Hz. Therefore, in order to make $J \ll \omega_0$, we should take an 
external electric field such that $\omega_0$ be at least as $\omega_0 \sim 10^{14}$ 
Hz. This corresponds to the mid-infrared range of electromagnetic spectrum, with the 
wavelength $\lambda \sim 10^{-3}$ cm. 

Keeping in mind ferroelectrics of the order-disorder type, with proton or deuteron 
bonds, we have the electric dipoles $d_0\sim e_0l_0$, in which 
$e_0=1.602177\times10^{-19}$ C is the proton charge and the length 
$l_0\sim 10^{-8}$ cm is the distance between the minima of the effective 
double-well potential. Since one Coulomb is 1 C$=2.997924\times 10^9$ g$^{1/2}$ 
cm$^{3/2}$ s$^{-1}$, then $e_0 = 4.803205 \times 10^{-10}$ g$^{1/2}$ cm$^{3/2}$ 
s$^{-1}$. Using the relation
$$
1{\rm D} = 10^{-18} {\rm erg}/{\rm G} = 
0.333564\times 10^{-27}{\rm C} \; {\rm cm} \; , 
$$
with $1 {\rm G}^2 = 1 {\rm erg}/{\rm cm}^3$, we find 
$$
d_0 \sim 10^{-18} {\rm erg}/{\rm G} \sim 1 {\rm D} \; .
$$
This gives the peak radiation intensities (\ref{104}) and (\ref{105}) at the moment 
of time $t_0$ of the order
\be
\label{106}
I_{inc}(\bfe_x,t_0) \sim N \times 10^{-20} \; {\rm W} \; , \qquad 
I_{coh}(\bfe_x,t_0) \sim N^2 F(k_0\bfe_x) \times 10^{-21} \;{\rm W} \; .
\ee

The number of atoms that certainly can radiate coherently is $N_{coh}\sim\rho\lbd^3$. 
With the average density $\rho\sim 10^{23}$ cm$^{-3}$ and the wavelength 
$\lbd\sim 10^{-3}$ cm, we get $N_{coh} \sim 10^{14}$. If the wavelength is larger 
than the sample length, then 
$$
F(k_0\bfe_x) \cong 1 \qquad ( \lbd > L ) \; .
$$
Therefore, we get
\be
\label{107}
I_{inc}(\bfe_x,t_0) \sim 10^{-6} \; {\rm W} \; , \qquad 
I_{coh}(\bfe_x,t_0) \sim 10^{8} \; {\rm W} \; .
\ee

If the sample has the cylindric shape, then the radiation beam, spreading along the 
cylinder axis, can split into filaments, each radiating coherently. This phenomenon 
of filamentation is well known experimentally (see, e.g., \cite{Sanz_63,Sanz_64,Leyva_65} 
and review articles \cite{Yukalov_48,Yukalov_66}) and has been explained theoretically 
\cite{Yukalov_40,Yukalov_67,Yukalov_68,Yukalov_69,Yukalov_70}. Each filament, radiating 
coherently, and having the volume $V_{coh} = \pi r_f^2 L$, contains the number of 
particles  
$$
N_{coh} = \rho V_{coh} = \pi \rho r_f^2 L \; .
$$
The filament radius is found \cite{Yukalov_40,Yukalov_67,Yukalov_68,Yukalov_69,Yukalov_70}
to be $r_f = 0.3 \sqrt{\lambda L}$, hence
$$
N_{coh} \approx 0.283 \rho \lbd L^2 \; .
$$
For the sample of length $L = 0.1$ cm, the number of coherently radiating atoms in a 
filament can reach $N \sim 10^{17}$. But when $\lambda \ll L$, it is necessary to take 
into account that for real samples, the scaling of the coherent radiation intensity with 
the number of atoms $N$ is not exactly $N^2$, but usually is lower, being caused by the 
geometric factor $F({\bf k})$ and different experimental imperfections \cite{Angerer_71}. 
For an estimate, we can take
$$
F(k_0\bfe_x) \sim \frac{\lbd^2}{\pi^2 L^2} \qquad ( \lbd \ll L ) \; .
$$  
Then, with the sample length $L = 0.1$ cm, we get  
$$
 I_{inc}(\bfe_x,t_0 ) \sim 10^{-3} \; {\rm W} \; , \qquad
 I_{coh}(\bfe_x,t_0 ) \sim 10^{10}\; {\rm W} \; ,  .
$$

\section{Conclusion}

A ferroelectric is inserted into a resonator cavity. The motion of ferroelectric 
dipoles induces in the cavity an electric field acting back on these dipoles. The 
ferroelectric is prepared in a nonequilibrium initial state, in an electric field 
directed opposite to ferroelectric dipoles. The motion of dipoles is triggered by 
the dipolar waves. The correlation between the moving dipoles is due to the resonator 
feedback field. The coherent motion of these electric dipoles produces coherent 
radiation, called superradiance.

The physics of the ferroelectric superradiance is principally different from atomic
superradiance. In the latter, the process is initiated by spontaneous atomic radiation, 
while in ferroelectric superradiance, the process is triggered by dipolar waves. In 
atomic superradiance, collective radiation is induced by effective atom interactions 
through the common electromagnetic radiation field. While in ferroelectric superradiance, 
the collectivization of dipole motion is induced by the resonator feedback field. No 
ferroelectric superradiance can develop without the resonator. The enhancement of 
radiation by a resonator is termed the Purcell effect \cite{Purcell_47}. But here, 
the resonator not merely enhances the coherent radiation, but induces coherence as 
such. Although mathematically there are direct analogies between spontaneous radiation 
of atoms and dipole waves and between the common radiation field of atoms and the 
resonator feedback field \cite{Yukalov_66,Yukalov_71}, but anyway the physics of 
atomic superradiance and ferroelectric superradiance is rather different.

Ferroelectric superradiance occurs in the infrared region. In that region, there 
exist many high-quality infrared resonators (see, e.g.,
\cite{Garin_72,Lecaplain_73,Osman_74,Radosavljevic_75,Yao_76,Xiao_77}). Therefore 
it looks quite straightforward to realize ferroelectric superradiance in experiment.

\section*{Acknowledgment}

I devote this paper to my dear friend V.V. Samartsev on the occasion of his 80-th 
jubilee. I have had a great pleasure of discussing with him a variety of problems 
related to the topic of the present paper.

\vskip 2mm  
Also, I am grateful for discussions and help to E.P. Yukalova.

\end{document}